\documentstyle[12pt]{article}
\input psfig.sty
\begin{document}

\def\vereq#1#2{\lower3pt\vbox{\baselineskip1.5pt \lineskip1.5pt
\ialign{$#1\hfill##\hfil$\crcr#2\crcr\sim\crcr}}}

\def\alt{\mathrel{\mathpalette\vereq<}}
\def\agt{\mathrel{\mathpalette\vereq>}}

\title{Anomalous Normal-State Properties of High-T$_c$ Superconductors --
Intrinsic Properties of Strongly Correlated Electron Systems?\\
Invited for publication in Advances in Physics}
\author{Th. Pruschke\\
Institut f\"ur Theoretische Physik, Universit\"at Regensburg,\\
93040 Regensburg, Germany,\\
M. Jarrell\\
Department of Physics, University of Cincinnati, \\
Cincinnati, Ohio, 45221,\\
and\\
J. K. Freericks\\
Department of Physics, Georgetown University\\
Washington, DC 20057}
\date{\today}
\maketitle
\begin{abstract}
A systematic study of optical and transport properties of the Hubbard model,
based on Metzner and Vollhardt's dynamical mean-field approximation,
is reviewed.   This model shows interesting anomalous properties
that are, in our opinion, ubiquitous to single-band strongly correlated systems
(for all spatial dimensions greater than one), and also compare qualitatively
with many anomalous transport features of the high-T$_c$ cuprates.
This anomalous behavior of the normal-state properties is traced to a
``collective single-band Kondo effect,'' in which a quasiparticle resonance
forms at the Fermi level as the temperature is lowered, ultimately yielding
a strongly renormalized Fermi liquid at zero temperature.
\end{abstract}

\newpage

\section{Introduction and survey}

\begin{figure}[htb]
{\centerline{\psfig{figure=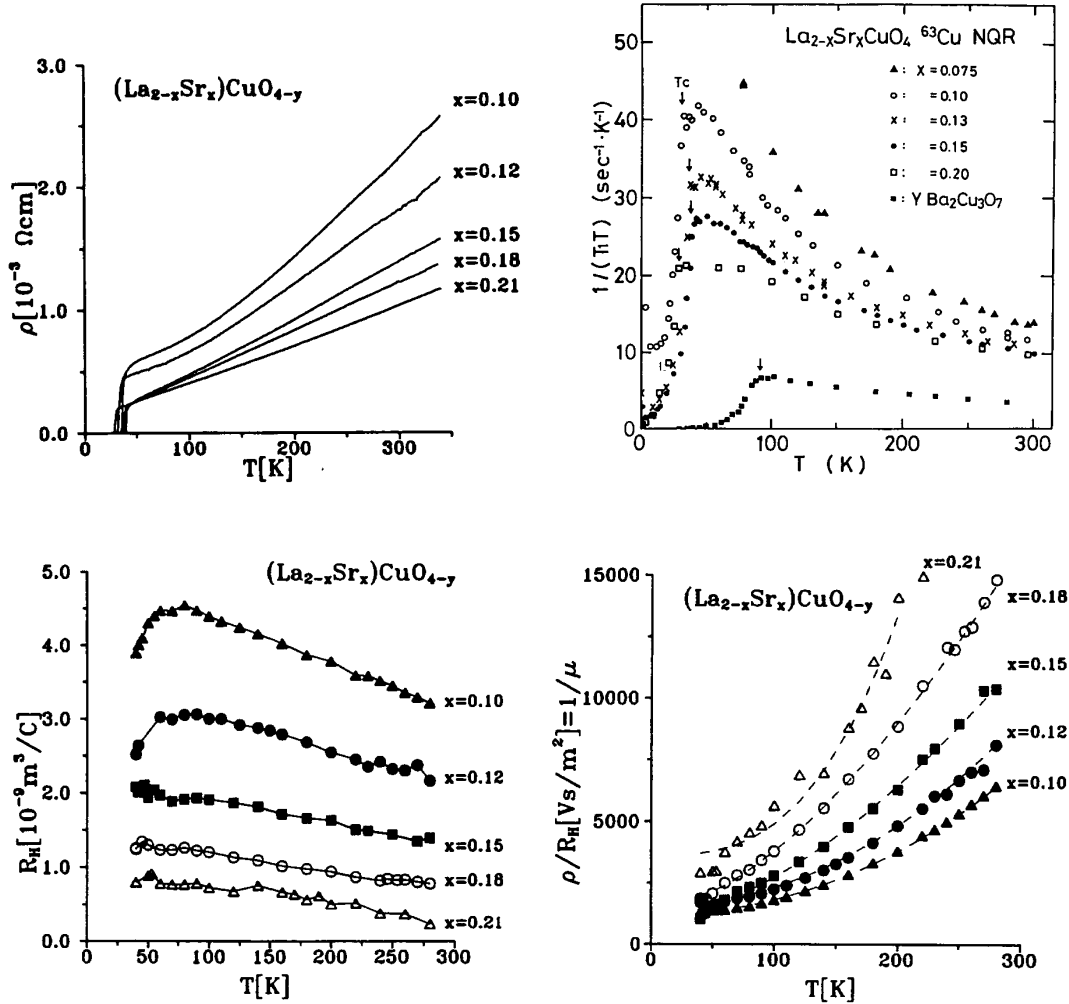,width=6.0in}}}

\caption[]{{\em Typical experimental
results for resistivity ($\rho$), NMR relaxation rate ($T_1$), Hall
coefficient ($R_H$), and inverse mobility ($1/\mu$) in high-$T_c$ compounds
(from \cite{christoph}).}}
\label{experiments}
\end{figure}

The discovery of the high-T$_c$ superconductors based on CuO-compounds
\cite{bedmull} has led to a large amount of theoretical work about the peculiar
properties of these materials. While initial work
concentrated on explaining ``why T$_c$ is so low'' and possible
exotic mechanisms for the superconductivity, it became obvious that an
understanding of the superconducting mechanism is linked to
\begin{figure}[htb]
{\centerline{\psfig{figure=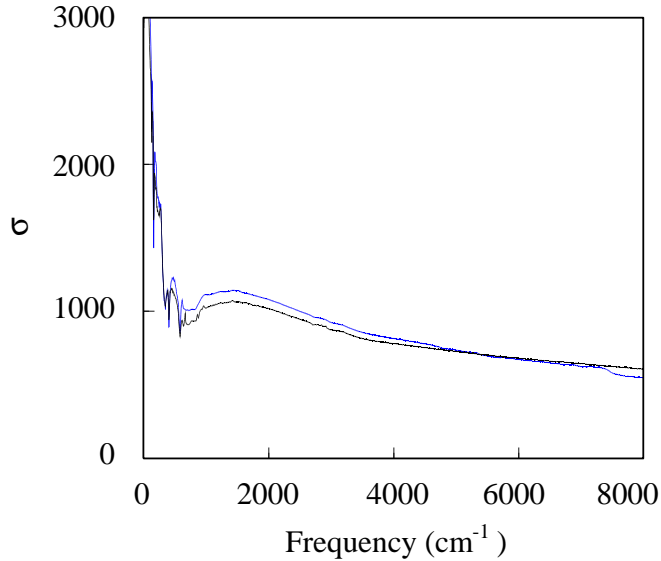,width=3.5in}}}
\caption[]{{\em Experimental optical conductivity of TlBa$_2$Ca$_2$Cu$_3$O$_9$
(from \cite{renk}).  Note the prominent mid-IR peak at about 1500 cm$^{-1}$.}}
\label{opt_exp}
\end{figure}
an understanding of the anomalous normal-state properties of these compounds
\cite{rew_oxides}. Most prominent among these are the linear  (in $T$)
resistivity, a linear (in $T$) NMR-relaxation rate
of the Cu-spins, and a Hall angle
that grows like $T^2$ over a rather wide temperature region
(see Fig.~\ref{experiments} \cite{christoph}).
Furthermore, the optical conductivity shows a Drude peak with a width
$1/\tau\sim T$ \cite{tanner_rev}, consistent with the linear
behavior of the resistivity and a pronounced mid-IR peak at
frequencies above the Drude peak (see Fig.~\ref{opt_exp} \cite{renk}).

Early on, it was argued \cite{anderson87} that most of these anomalous
properties can be explained by two special features appearing simultaneously
in these materials: (i) They are strongly correlated, i.e.\ their (effective)
local Coulomb interaction is comparable to or larger than the characteristic
kinetic energy of the relevant carriers, and (ii) they are highly anisotropic
with the electrons confined to the CuO-planes characteristic
for these compounds. This mixture of strong correlations and effective
2D-character is unusual and can only be found (apart from the
cuprates and other perovskites) in some Uranium-based heavy-fermion
compounds. The great interest in the high-temperature superconductors
has led to a number of interesting new
theoretical ansatzes, that, although based on the assumption of strongly
correlated carriers, focussed mainly on the 2D-character of the
CuO-planes. Most prominent are Anderson's concept of resonating valence-bonds
\cite{pwa_rvb}, his recent efforts to understand the electronic
properties phenomenologically in terms of a Luttinger-liquid fixed-point
(known from the theory of 1D electron systems) \cite{solyom,pwa_lutt}, Varma's
phenomenological marginal Fermi-liquid theory \cite{ffnj}, and Lee's
anyon concept emerging from a mapping of the 2D electron system onto a
nonlinear sigma-model \cite{anyon}. While the second and third ansatz have not
been ruled out, both the first and the last are insufficient
to account for the experimental data of these systems.

However, even in Anderson's and Varma's theories, one needs additional
assumptions about relaxation times \cite{pwa_lutt} or dynamical quantities
\cite{ffnj} to consistently account for {\em all}\/ anomalies.
Both theories also lack a thorough microscopic foundation. It is
unclear how the special form of the single-particle self
energy and the dynamical susceptibility (necessary for the marginal
Fermi-liquid picture) will emerge. Even introducing a special three-body
scattering vertex \cite{varma_three_body} does not resolve this difficulty,
since it merely shifts the problem to another level. A foundation for
Anderson's Luttinger-liquid picture is even less firm. The special
feature of 1D systems---that the exchange of charge over a given site is
strongly hindered by the Coulomb correlations, while the exchange of
spin costs almost no energy \cite{ogata_shiba} (which leads to the strict
separation of spin and charge degrees of freedom)---may already be
absent in 2D, since a physical electron may move ``around''
any other via a path in the plane.
Recent studies on the stability of the possible
low-energy fixed points in interacting fermion models also suggest that it
is the Fermi-liquid, rather than the Luttinger-liquid fixed point, that is
stable in 2D (unless the small-momentum-transfer couplings become
singular) \cite{castellani}.

There is another feature of strongly correlated systems that
is usually ignored in dealing with the cuprates. From the experimental and
theoretical investigation of heavy-fermion compounds it is known that the
strong correlations themselves already lead to anomalous features
in the magnetic and transport properties of metals. These systems are usually
three-dimensional in nature and, despite their complicated
bandstructure \cite{StiKub}, one can explain their physical properties
qualitatively (and to some extend even quantitatively) by merely taking into
account {\em one}\/ strongly correlated, localized band hybridizing with
{\em one}\/ uncorrelated conduction band \cite{gresteg}. This suggests
that the prominent anomalous features of the heavy-fermion materials are
intrinsic to the strong local correlations. Thus to unambiguously
understand the origin of any anomalous behavior in systems with strong
electronic correlations, it is necessary to discriminate between those
properties that are intrinsic and arise directly from the strong (local)
electron-electron scattering, and those that are connected to the geometry of
the underlying lattices, e.g.\ the 2D-character of the Cu-O planes
in the cuprates.

In previous publications \cite{pru_cox_ja,pru_ja} we demonstrated that even
for the simplest model of strong local correlations, the
single-band Hubbard model,
interesting anomalous results for several different transport quantities
occur which appear to be intrinsic properties of strongly correlated systems.
This latter statement is motivated by the fact that our calculations use
Metzner and Vollhardt's ``dynamical mean-field
approximation'' which may be viewed as the Fermionic equivalent of the
standard non-local mean-field description of models for magnetism. As is
well known, such a ``proper'' mean-field description will in general
reproduce the {\em intrinsic}\/ features of the underlying model.
Fluctuation corrections should only  affect phase
transitions and the quantitative values of temperature scales.

In this review we present a more detailed discussion of
the transport properties of strongly correlated systems described
by the Hubbard model in Eq.\ (\ref{hubmod}). Special emphasis
is made to thoroughly discuss the variation of quantities with
temperature and doping. We also try to make qualitative contact with recent
experiments on optical and transport properties of the cuprates. The paper
is organized as follows: In the next section an introduction into the
underlying theory (the so-called dynamical mean-field theory) is given. We also
give a brief account on how the transport properties are calculated. In
section 3 the systematics for several quantities as functions of temperature
and doping are presented and compared to experiment.
A discussion and outlook in section 4 conclude the review.

\section{Theoretical background}
The model discussed in this contribution is the
single-band Hubbard model \cite{hubbard}
\begin{equation}\label{hubmod}
H=\sum_{\vec{k}\sigma}\epsilon_{\vec{k}}c^\dagger_{\vec{k}\sigma}c_{\vec{k}\sigma}
+U\sum_in_{i\uparrow}n_{i\downarrow}\;\;.
\end{equation}
Our notation is the following:  $c^{\dagger}_{\vec{k}\sigma}$ is the creation
operator for an electron in a Bloch state with wavevector $\vec{k}$ and
$z$-component of spin $\sigma$, and $n_{i\sigma}$ is the electron number
operator for electrons localized at lattice site $i$ with spin $\sigma$
(i.e., the number operator for electrons in Wannier orbitals centered at
lattice site $i$).
For the sake of simplicity we restrict ourselves to a simple hypercubic
lattice (in $d$-dimensions)
with lattice constant $a$ and nearest-neighbor transfer $t$,
i.e.\ $\epsilon_{\vec{k}}= -2t\sum_{i=1}^d\cos(k_ia)$; $U$ denotes the local
Coulomb repulsion.  The energy scale is set by $4dt^2=:t^{*2}$ and the choice
$t^*=1$ (this is the required scaling for nontrivial results as $d\rightarrow
\infty$).

The Hubbard model is perhaps the simplest model of a correlated electronic
system, since it consists of a single band of delocalized electrons
subject to a {\em local}\/ Coulomb interaction. It is easier, therefore,  to
discriminate  intrinsic effects of the correlations (induced by the Coulomb
interaction) from bandstructure effects. Obviously, a more realistic
description of a condensed-matter system must take into account the
existence of more than one band, hybridizations between different bands,
and also additional local and long-range Coulomb interactions.
However, these more realistic models may be mapped onto
the Hubbard model (\ref{hubmod}) when excitations into these
other bands occur only virtually, leading to renormalized values
for e.g.\ $\epsilon_{\vec{k}}$ \cite{pru90}. Since the purpose of the
present contribution is to identify effects intrinsic to the local
correlations,
these other interactions may be neglected.

Models of interacting electrons on a lattice  are difficult to solve because
they include both {\em local} and {\em nonlocal} parts in the Hamiltonian.
The fundamental quantum-mechanical principle of
complementarity implies that one cannot expect to
describe such a system by either purely localized or purely delocalized
states. Approximations starting from either side (i.e.\
$\mbox{bandwidth}/U\ll1$
or $\mbox{bandwidth}/U\gg1$) are not generally
applicable to the intermediate regime
($\mbox{bandwidth}\approx U$). To obtain sensible results
for the Hubbard model in this intermediate regime, the Hamiltonian
(\ref{hubmod}) is usually studied
with exact diagonalization techniques \cite{dagotto} or quantum Monte Carlo
methods (QMC) \cite{hirsch}. Both methods share the problem that they
can only be applied to comparatively small system sizes, ($\sim 20$ sites in
exact diagonalization and $\sim 100$ sites in QMC [for not too low
temperatures]).  This restriction seriously affects any attempt
to determine low-energy/low-temperature properties and
makes the calculation of dynamical quantities and transport properties
difficult. An ansatz is therefore needed to calculate different physical
quantities of strongly correlated systems that (i) maintains the important
local correlations and (ii) allows calculations to be performed
in the thermodynamic limit.

Such a method was proposed by Metzner and Vollhardt \cite{mevoll} and
M\"uller-Hartmann \cite{muha89} who observed that
the renormalizations due to local two-particle interactions like the
Hubbard-$U$
become purely local as the coordination number of the lattice increases.
More precisely, the irreducible single-particle self energy
$\Sigma_{\vec{k}}(z)$ and the irreducible two-particle self energy
$\Gamma_{\vec{k},\vec{k}+\vec{q}}(z,z')$ both become independent of momentum
for large coordination number ($2d\to\infty$)
\cite{muha89,bramiel90}:
\begin{equation}
\begin{array}{l@{\ =\ }l}
\lim\limits_{d\rightarrow\infty}
\displaystyle\Sigma_{\vec{k}}(z) & \displaystyle\Sigma(z)\quad ,\\[5mm]
\lim\limits_{d\rightarrow\infty}
\displaystyle\Gamma_{\vec{k},\vec{k}+\vec{q}}(z,z') &
\displaystyle\Gamma(z,z')\;\;.
\end{array}
\label{loc_selfenergy}
\end{equation}
One can use standard techniques of field theory to show that the second
relation in Eq.\ (\ref{loc_selfenergy}) is a necessary
consequence of the first for {\em all} two-particle self
energies \cite{bramiel90,comment}.
A further consequence of the first relation is that the solution of the Hubbard
model may be mapped onto the solution of a local correlated system coupled
to an effective bath that is self-consistently determined by the following
five-step process \cite{bramiel90,janis,kimkura,volljan,jarrell91,georges92}:
\begin{itemize}
\item{(i)} Choose a suitable starting guess for the local self energy
$\Sigma(z)$.
\item{(ii)} Calculate the local single-particle Green function from its Fourier
transform
\begin{equation}
G_{ii}(z)=\frac{1}{N}\sum_{\vec{k}}\frac{1}{[G^{(0)}_{\vec{k}}(z)]^{-1}-\Sigma(z)}\;\;,
\end{equation}
with $G_{\vec{k}}^{(0)}:=1/(z+\mu-\epsilon_{\vec{k}})$ the {\it noninteracting}
momentum-dependent Green function.
\item{(iii)} Obtain the effective medium, denoted by ${\cal G}(z)$, by
subtracting off the local correlations from the Coulomb interaction
at site $i$,
\begin{equation}
[{\cal G}(z)]^{-1}:=(G_{ii})^{-1}+\Sigma(z)\;\;.
\end{equation}
\item{(iv)} Solve the local impurity problem defined by the effective medium
${\cal G}(z)$ and the local Coulomb interaction $Un_{i\uparrow}n_{i\downarrow}$
at site $i$ (via an exact quantum Monte Carlo algorithm, or an approximate
algorithm based upon a diagrammatic analysis)
to obtain a new local single-particle Green function $\tilde{G}(z)$.
\item{(v)} Obtain a new self energy via $\Sigma(z)=[\tilde{G}(z)]^{-1}-[{\cal
G}(z)]^{-1}$
and repeat steps (ii) through (v) until the local Green functions are
identical between two successive iterations $[G_{ii}(z)=\tilde{G}(z)]$.
\end{itemize}
This mapping onto a local problem coupled to an effective bath is reminiscent
of the standard mean-field description of spin systems. As is well
known, such a theory emerges systematically from the same limit of
large coordination number
\cite{itzykson}. One major difference between electronic models and
spin models is that in the former the Coulomb interaction introduces
a nontrivial local dynamic that is preserved in the
mean-field approach. The molecular field for the electronic mean-field
theory is not a constant number, but is instead a function of the energy,
hence the name ``dynamical mean-field theory''.  These local
electronic mean-field theories have a rich history, dating back to
Migdal and Eliashberg's theory of superconductivity \cite{migdal_eliashberg};
Metzner and Vollhardt \cite{mevoll} were the first to realize that
these mean-field theories become exact solutions in the limit of infinite
dimensions.

Although the solution of the effective local system is still a highly
nontrivial matter, it has the great advantage that
one is able to work directly in the thermodynamic
limit.  Note that the lattice structure has not been completely eliminated from
the problem: It enters indirectly via the free density of states (DOS),
when  momentum summations are converted to energy integrals, and it enters
in the evaluation of susceptibilities, which still maintain a (weak)
momentum dependence.

Two methods have proven to be most successful in solving the remaining
local problem:
a QMC scheme based on the work of Hirsch and Fye \cite{hifye,jarscal,jarrell91}
and a perturbational method known as the non-crossing
approximation (NCA) \cite{nca}. Both
methods have their merits and disadvantages and may be viewed as complementary
to each other, providing a means of accessing many different
quantities of interest. A more detailed discussion of the limitations
of both approaches can be found in our earlier
publications \cite{pru_cox_ja,japru}.

\begin{figure}[htb]
{\centerline{\psfig{figure=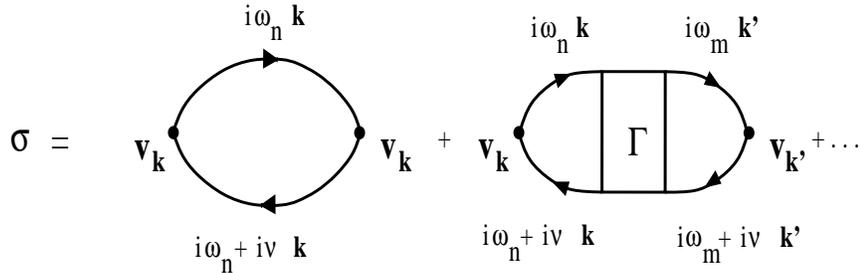,width=4.5in}}}

\caption[]{{\em Diagrammatic representation of the first two contributions
to the conductivity. The second diagram contains a {\em full}\/
particle-hole vertex insertion. The latter is momentum independent,
i.e.\ the $\vec{k}$-sums on both sides can be performed independently.
Since the current vertex and the single-particle Green functions are of
different parity (with respect to their $\vec{k}$-dependence) the
second and all higher-order diagrams identically vanish.}}
\label{conddiagr}
\end{figure}

The knowledge of the single-particle self energy and the
{\em local}\/ two-particle vertex functions (which can also be
determined from the same local problem) enables one to calculate physical
quantities. As an illustration of this, an extended discussion of magnetism in
the Hubbard model may be found in Refs.\ \cite{japru,freeja,pvd}. Other
quantities
of interest include the transport coefficients, the optical conductivity, the
thermopower,
and the Hall coefficient. For example, the conductivity can be calculated
exactly in the dynamical mean-field theory by the following procedure:
Figure~\ref{conddiagr} shows the leading diagrams in the
expansion of the conductivity. As was discussed earlier, the particle-hole
vertex appearing in the second and all higher-order diagrams is
momentum independent. This means that the $\vec{k}$-sums
on the left and right end of all diagrams, except for the simple bubble,
may be performed independently.  Since the current operator contains
the $\vec{k}$-gradient of the kinetic energy, and the Green functions are
$\vec{k}$-dependent only through $\epsilon_{\vec{k}}$, these sums identically
vanish. Thus the conductivity is given by the simple bubble
only \cite{pru_cox_ja,khurana}. The evaluation of the bubble leads to
\begin{equation}\label{optcond}
\sigma_{xx}(\omega) = \frac{\pi e^2}{2\hbar a}\int\limits_{-\infty}^\infty
d\epsilon
\frac{f(\epsilon)
-f(\epsilon+\omega)}{\omega}\frac{1}{N}
\sum_{\vec{k}\sigma}\left(\frac{\partial\epsilon_{\vec{k}}}{\partial
k_x}\right)^2
A_{\vec{k}}(\epsilon)A_{\vec{k}}(\epsilon+\omega)\;\;,
\end{equation}
which reduces to
\begin{equation}\label{dccond}
\sigma = \frac{e^2\pi}{2\hbar a}\int\limits_{-\infty}^\infty d\epsilon
\left(-\frac{\partial f(\epsilon)}{\partial\epsilon}\right)\frac{1}{N}
\sum_{\vec{k}\sigma}\left(\frac{\partial\epsilon_{\vec{k}}}{\partial
k_x}\right)^2
[A_{\vec{k}}(\epsilon)]^2\;\;,
\end{equation}
for the dc conductivity, where $a$ is the lattice constant, the spectral
weight satisfies
$A_{\vec{k}}(\omega):=-\frac{1}{\pi}{\rm{Im}}\left[G_{\vec{k}}(\omega)\right]$,
and $f(\epsilon ):=1/[1+\exp\beta\epsilon ]$ is Fermi's function. With
$h/e^2\approx2.6\cdot 10^{4}\Omega$ the constants in front of
Eq.\ (\ref{dccond}) can be evaluated to yield $\sigma_0 \sim
10^{-3}\ldots 10^{-2} \left[(\mu\Omega cm)^{-1}\right]$.

The Drude weight $D$, may be determined by extrapolation of the
Matsubara-frequency current-current correlation function using
the method proposed by Scalapino et al.\cite{scal_white_zhang}.
This method sets a criteria to determine if the
ground state of a system is a metal, insulator, or superconductor, by
determining the asymptotic form of the current-current susceptibility in
the x-direction, $\Lambda_{xx}({\bf{q}},i\nu_n)$ where
$\nu_n=2n\pi T$.  Specifically, the Drude weight, $D$, is given by
\begin{equation}\label{drude_weight}
D=\pi\lim_{T\to 0}\left[\left< -T_x \right>e^2 - \Lambda_{xx}({\bf{q}}=0,
2\pi iT)\right]\,.
\end{equation}
Here, $\left< -T_x \right>$ is the average kinetic energy per site, divided by
the number of lattice dimensions.

Comparison of Eq.~(\ref{optcond})  with the standard expressions for transport
coefficients \cite{madelung} in the relaxation-time approximation, shows
that a variety of other transport coefficients may be calculated
if one identifies
\begin{equation}\label{tauxx}
\tau_{xx}(\epsilon) := \frac{1}{N}\sum_{\vec{k}\sigma}
\left(\frac{\partial\epsilon_{\vec{k}}}{\partial k_x}\right)^2
 [A_{\vec{k}}(\epsilon)]^2
\end{equation}
as the transport relaxation time. For example,
the electronic contribution to the thermopower becomes \cite{madelung}
\begin{equation}\label{thermopower}
S=-\frac{k_{\rm B}}{|e|}\beta\frac{L_{12}}{L_{11}}\;\;,\frac{k_{\rm B}}{|e|}
\approx 86\left[\mu V/K\right]\quad ,
\end{equation}
where $L_{jk}$ are the
standard transport integrals\begin{equation}\label{transportintegrals}
L_{jk}=\int\limits_{-\infty}^\infty d\epsilon\left(-\frac{\partial
f(\epsilon)}{\partial\epsilon}\right)\tau^j(\epsilon)\epsilon^{k-1}\quad .
\end{equation}

It is known from studies of heavy fermion systems that the relaxation-time
approximation is sufficient to understand most zero-field properties. The Hall
coefficient, however, is more sensitive to the approximations involved:
Within the standard relaxation-time approximation the Hall coefficient
satisfies
\begin{equation}\label{hall_coeff}
R_H=\frac{1}{-|e|}\frac{L_{11}}{L_{21}}\;\;.
\end{equation}
Note that both $L_{11}$ and $L_{21}$ are positive, implying that the
Hall-coefficient is always negative and that the transport is electronlike.
Experiments, on the other hand, show that $R_H$ can change sign
in intermediate or low temperature regions, and that the Hall coefficient
is usually positive with an anomalous temperature dependence.
An explanation for this theoretical deficiency
was proposed by the phenomenological introduction of a
``skew-scattering'' term \cite{coleman}. However, a more refined
treatment of the field-dependent conductivity can be performed
\cite{vuruganti}.  The result is
\begin{equation}\label{hallcond}
\sigma_{xy}^H=
\frac{2\pi^2|e|^3aB}{3\hbar^2}\int\limits_{-\infty}^\infty d\omega
\left(\frac{\partial f(\omega)}{\partial\omega}\right)
\frac{1}{N}\sum_{\vec{k}\sigma}\left(\frac{\partial\epsilon_{\vec{k}}}{\partial
k_x}\right)^2\frac{\partial^2\epsilon_{\vec{k}}}{\partial k_y^2}
[A_{\vec{k}}(\omega)]^3\,,
\end{equation}
for the Hall conductivity. Here, $B$ is the external magnetic
field which points in the $z$-direction. The constants in
Eq.\ (\ref{hallcond}) can be
rearranged according to ${|e|^3aB}/{\hbar^2}=\sigma_0{|e|a^2B}/{\hbar}$
and ${|e|a^2B}/{\hbar}\approx 10^{-5}B\left[\frac{1}{T}\right]$.
Inserting the values for $\sigma_0$, we obtain
$R_H={\sigma_{xy}}/{\sigma_{xx}^2}\sim10^{-8}\ldots 10^{-9}B\left[{m^3}/{cT}
\right]$ as the unit for the Hall coefficient.

In the limit of large coordination number, and
for a simple hypercubic lattice,
the $\vec{k}$-sums in Eqs. (\ref{tauxx}) and (\ref{hallcond})
can be further simplified to yield
\begin{equation}\label{int_ident1}
\frac{1}{N}\sum_{\vec{k}\sigma}
\left(\frac{\partial\epsilon_{\vec{k}}}{\partial k_x}\right)^2
K(\epsilon_{\vec{k}}) = \frac{2}{d}\int\limits_{-\infty}^{\infty} d\epsilon
N_0(\epsilon)K(\epsilon)
\end{equation}
and
\begin{equation}\label{int_ident2}
\frac{1}{N}\sum_{\vec{k}\sigma}\left(\frac{\partial\epsilon_{\vec{k}}}{\partial
k_x}\right)^2\frac{\partial^2\epsilon_{\vec{k}}}{\partial k_y^2}
K(\epsilon_{\vec{k}})= -\frac{1}{2d^2}\int\limits_{-\infty}^{\infty} d\epsilon
N_0(\epsilon)\epsilon K(\epsilon)\;\;,
\end{equation}
with $K(\epsilon_{\vec{k}})$ an arbitrary function depending
on $\vec{k}$ through $\epsilon_{\vec{k}}$ only, and
$N_0(\epsilon ):=e^{-(\epsilon/t^\ast)^2}/(\sqrt{\pi}t^\ast)$ the
noninteracting DOS.

Examination of Eqs.~\ref{optcond},\ref{hallcond}--\ref{int_ident2} shows that
the electronic conductivity is a
$1/d$ effect and that the Hall effect enters to order $1/d^2$. Note, however,
that $R_H=\sigma_{xy}/\sigma_{xx}^2$ enters to zeroth order!

Another important probe of the high-temperature superconductors involves
nuclear magnetic resonance (NMR).  The spin-lattice relaxation time $T_1$
characterizes the time needed to align the nuclear spins along the direction
of the field.  The local NMR-relaxation rate satisfies
\begin{equation}\label{t1def}
\frac{1}{T_1}=T\lim_{\omega\rightarrow 0} {\rm Im} \frac{\chi(\omega)}{\omega}
\quad ,
\end{equation}
where $\chi(\omega)$ is the local dynamic spin susceptibility.  Similarly,
the spin-echo decay rate (or transverse nuclear relaxation rate) $T_{2G}$ is
another probe of the spin dynamics in the cuprates.  $T_{2G}$ satisfies a more
complicated relation \cite{penn_slich,thelen_pines}
\begin{equation}\label{t2def}
\left [ \frac{1}{T_{2G}}\right ]^2 = C \left [
\sum_{\vec{k}}F(\vec{k})^4\chi(\vec{k})^2
- \{\sum_{\vec{k}}F(\vec{k})^2\chi(\vec{k})\}^2\right ] \quad ,
\end{equation}
with $\chi(\vec{k})$ the static momentum-dependent spin susceptibility, $C$ an
overall normalization factor,
and $F(\vec{k})$ the relevant form factor.  The form factor involves the local
site and the nearest-neighbor shell, and assumes the form
\begin{equation}\label{formfac}
F(\vec{k})=1+\gamma \frac{2t}{U}\epsilon_{\vec{k}}\quad ,
\end{equation}
in infinite dimensions.  Here, $\gamma$ is a constant of order 1.  The static
susceptibility turns out to be a function of $\epsilon_{\vec{k}}/\sqrt{d}=:X$
(in infinite dimensions), with $X=0$ corresponding to the local susceptibility.
The integrals over the form factors can be performed, to yield
\begin{equation}
\label{t2gfinal}
\left [ \frac{1}{T_{2G}}\right ]^2 = \frac{C}{d} \left [ 2\gamma^2\chi^2(X=0)-
\frac{1}{2} \Big | \frac{d\chi(X)}{dX}\Big |^2_{X=0}\right ] \quad .
\end{equation}
The inverse spin-echo decay rate is a $1/\sqrt{d}$ effect in infinite
dimensions,
and it is effectively proportional to the local susceptibility, since the
derivative of the momentum-dependent susceptibility with respect to $X$ is
an order of magnitude smaller than the susceptibility itself in the region
of interest (see, for example Fig. 1 of \cite{freeja}).
In fact, we neglect the derivative term when evaluating $T_{2G}$
[$T_{2G}\propto
1/\chi(X=0)$].

\section{Discussion of the results}

\subsection{Single-particle properties}
\paragraph{Theory.}
Before we present results for the transport properties, we will summarize
some of the basic physics in the Hubbard model that is obtained from the
dynamical mean-field theory.
The Coulomb parameter $U$ is fixed at $U=4$. This choice may appear to be
arbitrary at a first glance, however, the main effect of $U$ in the
strong-coupling regime is to determine the characteristic low-energy
scale \cite{japru}, and $U=4$ is a convenient choice
for numerical and presentational reasons.

As the temperature is lowered, the Hubbard model in infinite-d is found
to always be a Fermi liquid\cite{jarrell91,pru_cox_ja}, except
for the region of phase
space where it is magnetic\cite{freeja,pvd}.  A Fermi liquid is defined by a
self-energy that has the following structure:
\begin{eqnarray}
{\rm Re}\Sigma(\omega+i0^+)&=&{\rm Re}\Sigma(0)+\omega (1-Z)+O(\omega^2)
\quad ,\cr
{\rm Im}\Sigma(\omega+i0^+)&=&-\Gamma+O(\omega^2)\quad ,
\label{eq: flt}
\end{eqnarray}
with $\Gamma\propto T^2$ for temperatures $T\ll T_0$ the characteristic
low-temperature scale. [The Fermi temperature $T_0$ decreases to zero
as half filling is approached\cite{pru_ja}, and the Fermi-liquid-theory form of
Eq.~(\ref{eq: flt}) {\it still holds for moderate temperatures},
with the only change
being $\Gamma\propto T$ for $T>T_0$.] The spectral weight, then assumes the
form
\begin{equation}
A_{\vec{k}}(\omega) = {1\over\pi} {\Gamma\over \Gamma^2+(\omega
Z+\epsilon_F-\epsilon_{\vec{k}})^2}+
A^{Inc}_{\vec{k}}(\omega)\quad ,
\label{eq: spec_fun_flt}
\end{equation}
with the Fermi level defined by $\epsilon_F := \mu - {\rm Re}\Sigma(0)$ and
$A^{Inc}_{\vec{k}}(\omega)$ denoting the (rather structureless) incoherent
contributions
to the spectral function.  The spectral function includes a delta function
at zero temperature [$A_{\vec{k}}(\omega) \rightarrow \delta(\omega
Z+\epsilon_F-\epsilon_{\vec{k}})+
A^{Inc}_{\vec{k}}(\omega)$] because the broadening $\Gamma$ vanishes in that
limit.  The single-particle DOS, $N(\omega)$, is defined to be the integral
of the spectral function over all momentum,
\begin{equation}\label{dosdef}
N(\omega):=\sum_{\vec{k}}A_{\vec{k}}(\omega)\quad .
\end{equation}

\begin{figure}[htb]
{\centerline{\psfig{figure=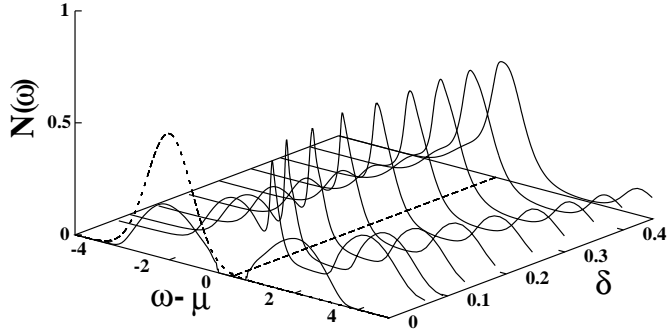,width=3.5in}}}
\caption[]{{\em Single-particle DOS of the Hubbard model for different dopings
$\delta=1-\langle n\rangle$ at an inverse temperature $\beta=43.2$. At half
filling ($\delta=0$)
the DOS has a gap between the lower and upper Hubbard band. Away from half
filling a resonance occurs at the chemical potential in addition to the
lower and upper Hubbard bands. This resonance becomes broader with
increasing $\delta$ and finally merges with the lower Hubbard band.
The dotted curve at $\delta=0$ displays the (shifted) noninteracting DOS.}}
\label{DOSdope}
\end{figure}

Figure~\ref{DOSdope} shows the evolution of the single-particle
density of states as a function of doping for a fixed temperature
($\beta=43.2)$.
In addition, the (shifted) DOS for the noninteracting system has been added
for comparison (dotted line at $\delta=0$).
At half filling, the doping satisfies $\delta=0$, and the system shows a
pseudogap in the DOS.
Note that the lower and upper Hubbard bands centered at $\omega
\approx \pm U/2$ have the same approximate width as the free DOS, but are
decreased by a factor of two from the unperturbed height due to the
correlations.  Away from half filling ($\delta\ne 0$), a sharp resonance
appears near the chemical potential. As the doping $\delta$ increases, the
width
of this resonance also increases and it starts to merge with
the lower Hubbard band.  For dopings $\delta>0.4$ both low-energy
peaks become indistinguishable, implying that the system has become
an uncorrelated metal.  The upper Hubbard band
seen in Fig.~\ref{DOSdope} is, on the other hand, well separated from
the low-energy excitations by a pseudogap of order $U$ and thus contributes
only to high-energy features in e.g.\ the optical conductivity
\cite{pru_cox_ja}.

\begin{figure}[htb]
{\centerline{\psfig{figure=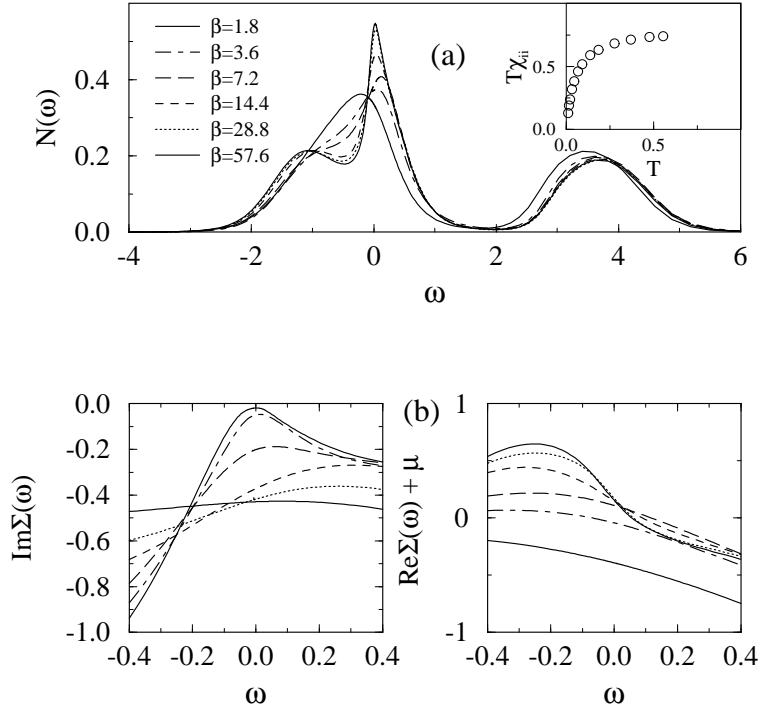,width=3.5in}}}

\caption[]{(a) {\em The evolution of the density of states when $U=4$
and $\delta=0.188$. The development of a sharp peak at the Fermi
surface is correlated with the reduction of the screened local moment
$T\chi_{ii}(T)$, as shown in the inset. Hence the development of the
peak may be associated with a resonant Kondo screening of the spins.}\\
(b) {\em The real and imaginary parts of the self energy for various
temperatures when $U=4$ and $\delta=0.188$.  Note that as the temperature is
lowered, $\rm{Im} \Sigma(\omega)$ becomes parabolic in $\omega$, indicating
the formation of a Fermi liquid.}}
\label{DOStemp}
\end{figure}

\begin{figure}[htb]
{\centerline{\psfig{figure=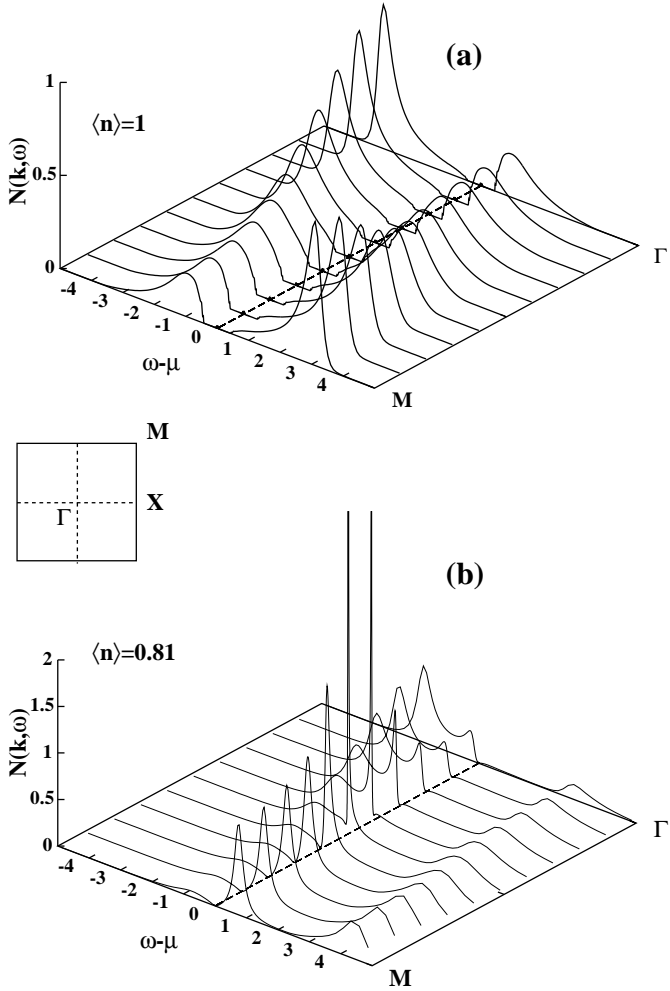,width=3.5in}}}

\caption[]{\em
(a) Angle resolved spectra for $U=4$, $\beta=43.2$ and $\langle n\rangle=1$ (a)
and $\langle n\rangle=0.8122$ (b)
along the $\Gamma$--$M$ direction of a two-dimensional
Brillouin zone. As expected at half filling, the system shows a gap at $\mu$
throughout
the entire zone. Away from half filling a quasi-particle peak develops when
the Fermi surface is crossed. In addition, there are strongly damped features
at the position of the upper Hubbard band. The data for half filling were
obtained with the NCA.}
\label{dosofk}
\end{figure}
\begin{figure}[htb]
{\centerline{\psfig{figure=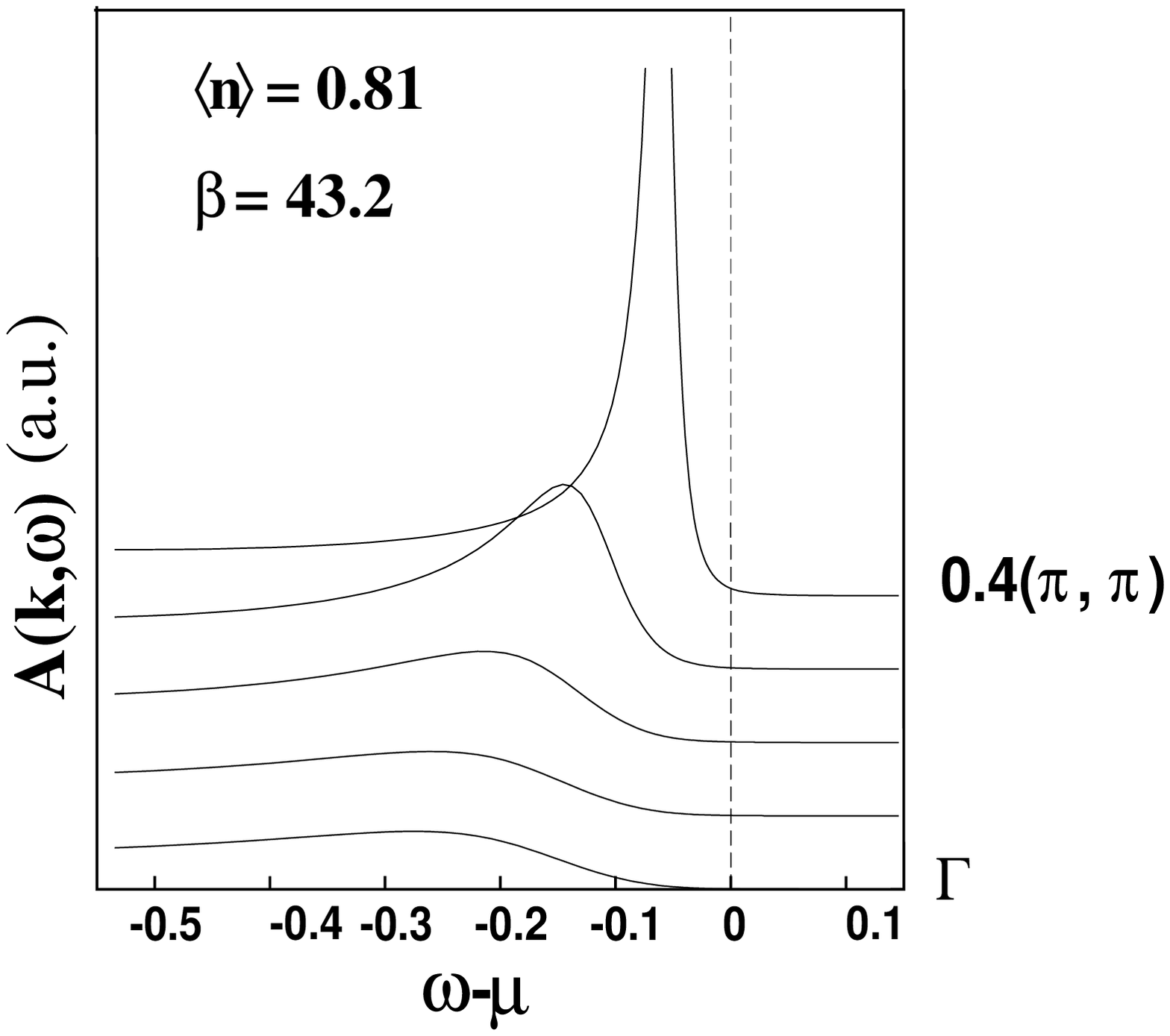,width=3.5in}}}

\caption[]{\em
The data in Fig.~\ref{dosofk}b for $\Gamma$--$M$ multiplied by Fermi's function
(at $\beta=43.2$) and restricted to the low-energy part. This is the type of
spectrum that would typically be observed in an ARPES
experiment (see e.g. \cite{olson}). Data for $\vec{k}$ beyond $0.4(\pi,\pi)$
were dropped because these features are cut off by Fermi's function.}
\label{arpes}
\end{figure}
Figure~\ref{DOStemp} collects the behavior of the DOS (Fig.~\ref{DOStemp}a)
and the real and
imaginary parts of the single-particle self energy (Fig.~\ref{DOStemp}b)
as the temperature is reduced at fixed doping ($\delta=0.188$).
The resonance at $\mu$ appears to be strongly temperature dependent, vanishing
as $T$ increases. For low temperatures, $T\ll T_0$,
the peak is centered near the chemical potential, and the height saturates
to the value of the noninteracting DOS
$N_0(\omega=0)=1/\sqrt{\pi}\approx 0.56$.  The upper bound for the local DOS
follows from the following simple argument (that does not require
Fermi-liquid-theory behavior): Since
the self energy is $\vec{k}$-independent, the interacting DOS may be written as
\begin{equation}
N(\omega)=\int\limits_{-\infty}^\infty d\epsilon N_0(\epsilon)
\left(-\frac{1}{\pi}\right){\rm Im}\frac{1}{\zeta-\epsilon}
\end{equation}
with $\zeta:=\omega-\Sigma(\omega+i0^+)$ being a complex number. Applying
the mean-value theorem, we obtain
\begin{equation}\label{dos_sumrule}
N(\omega)=N_0(\xi)\int\limits_{-\infty}^\infty d\epsilon
\left(-\frac{1}{\pi}\right){\rm Im}\frac{1}{\zeta-\epsilon}=N_0(\xi)\le N_0(0)
\end{equation}
with a suitably chosen $\xi$. This implies that we must always have
$N(\omega)\le N_0(0)={1}/{\sqrt{\pi}}$, but it {\em does not}\/ tell us
whether and where this maximal value will be reached (although the maximum is
attained at $\omega=-\epsilon_F/Z$ as $T\rightarrow 0$ for a Fermi liquid).
Since, on the other hand, the local
DOS is obtained from an effective single-impurity Anderson model, we expect
strong resonant scattering at $\mu$ in the correlated limit, i.e.\ the
system tends to provide a large DOS near the Fermi level.
However,
it cannot build up DOS everywhere, but has to adjust itself in such a way that
the sum rule in Eq.\ (\ref{dos_sumrule}) is fulfilled. The system, therefore,
satisfies a delicate self-consistent Friedel's sum rule.


\begin{table}[htb]
\begin{center}
\begin{tabular}{|c|c|} \hline
$\delta=1-\langle n\rangle   $& $T_0$ \\ \hline\hline
$0.0680$& 0.0177 \\
$0.0928$& 0.0273 \\
$0.1358$& 0.0478 \\
$0.1878$& 0.0730 \\
$0.2455$& 0.1074 \\ \hline\hline
\end{tabular}
\caption[]{\em{Values of the Kondo scale $T_0$ for different dopings
when $U=4$.}}
\end{center}
\end{table}

The development of the observed resonance
as $T\to 0$ is accompanied by a strong reduction of the effective local
magnetic moment $T\chi_{ii}(T)$ (see inset to Fig.~\ref{DOStemp}a).
\begin{figure}[htb]{\centerline{\psfig{figure=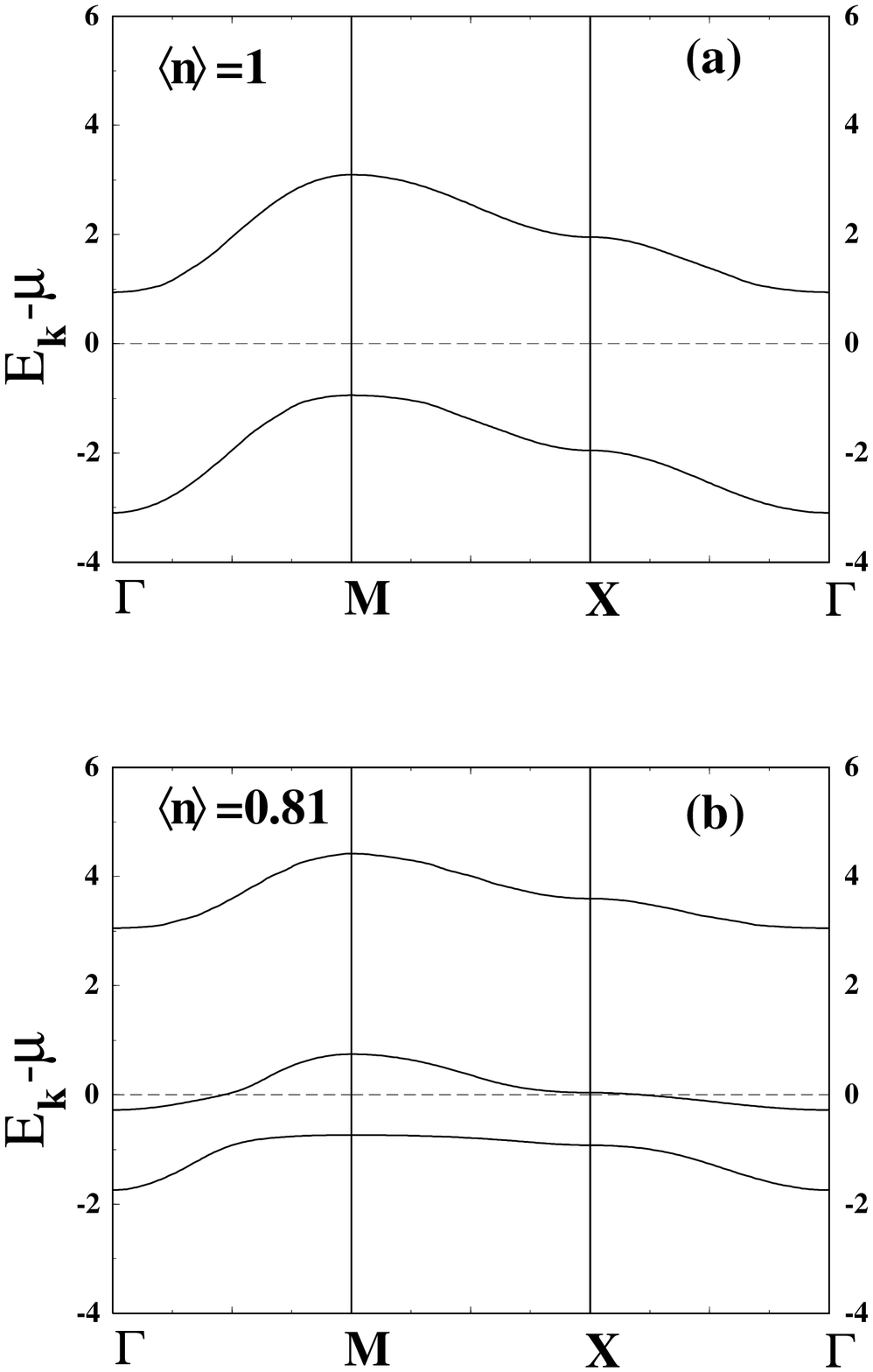,width=3.5in}}}

\caption[]{\em Bandstructure for the Hubbard
model obtained from the self-energy in Fig.~\ref{dosofk}.}
\label{bandstruct}
\end{figure}
This simultaneous appearance of the resonance at $\mu$ and vanishing of the
effective local moment suggests that this effect may be attributed to a
Kondo-like quenching
of the local magnetic moment in the system \cite{pru_ja,japru}.
The corresponding low-energy scale, which also sets the width of the resonance
at
$\mu$ seen in Figs.~\ref{DOSdope} and \ref{DOStemp}, may be estimated from
$1/\chi_{ii}(T=0)$ and is tabulated in Table 1.
Note, however,
that unlike conventional Kondo systems described by Anderson's model, where
one uncorrelated band couples to a localized correlated state, leading to the
quenching of the latter's moments,
there exists only one type of electron in the Hubbard model (\ref{hubmod})
which provides the band that is used to quench its own moments.
This is a new effect which we propose to be called
the ``collective single-band Kondo effect''.

We can also examine the angle-resolved photoemission spectra (ARPES) obtained
from the self-energy. Although it is a crude approximation, we assume
that the local self energy describes the self energy of
a two-dimensional Hubbard model on a square lattice.
Taking a cut along the $\Gamma$--$M$ direction in the Brillouin zone we
obtain the spectra shown in Fig.~\ref{dosofk} for $U=4$, $\beta=43.2$,
$\langle n\rangle=1$ (Fig.~\ref{dosofk}a) and
$\langle n\rangle=0.8122$ (Fig.~\ref{dosofk}b) (note that this approximation
does not distinguish between a cut along the diagonal $\Gamma-M$ or along the
$k_x$ and $k_y$ axes $\Gamma-X-M$ because the self energy is local, and does
not depend upon $\vec{k}$). While the system is clearly
insulating and non Fermi-liquid like at half filling, a typical quasiparticle
peak appears when the Fermi surface is crossed away from
half filling. In addition two strongly damped bands (the lower and upper
Hubbard bands) can be seen at higher energies in both figures.
When the data of Fig.~\ref{dosofk}b  is multiplied by Fermi's function (at
$\beta=43.2$),
one arrives at the typical ARPES result of Fig.~\ref{arpes}.
This is quite analogous to what is seen in
QMC simulations for true 2D-clusters \cite{bulut}.

Using the peak positions in Fig.~\ref{dosofk} as a definition of the
quasiparticle energies $E_{\vec{k}}$, yields the bandstructure shown in
Fig.~\ref{bandstruct}.
At half filling (Fig.~\ref{bandstruct}a) there are two cosine-like bands
below and above $\mu$, which represent the lower and upper Hubbard bands.
They are separated by an indirect gap of order $U/2$.
Off half filling, the upper and lower Hubbard bands are flattened relative to
their values at half filling (with the most flattening occurring near
the Fermi level).  In addition, a dynamically generated flat quasiparticle band
can also be seen.  This band is the flattest of the three and has
surprisingly small dispersion near the $X$ point,
which compares qualitatively to those obtained for small two-dimensional
Hubbard-clusters \cite{bulut,preuss}, and supports the
conjecture that the dynamical mean-field theory already gives
an accurate picture for the single-particle properties down to $d=2$.
Note that the flatness near the $X$ point arises from both the flatness of the
noninteracting bandstructure, and the many-body renormalizations that reduce
the quasiparticle bandwidth.

\paragraph{Experiment.}
Only qualitative comparisons can be made between experimental photoemission
spectroscopy (PES) and inverse photoemission spectroscopy (IPES) and
theoretically generated spectra, because the experimental resolution
broadening, selection rules for the matrix elements, and the internal
dynamics of the scattering and relaxation processes, ultimately alter the
experimentally measured spectrum from its theoretical counterpart.
In addition (with the exception of first-principles calculations), simplified
electronic models are used, that are  restricted to specific energy regions
(mostly the low-energy ones). Nevertheless one may at least identify certain
features of the model with trends found in experiments.

The measured PES and IPES spectra of the cuprates have as their most
interesting
trend, an increase of spectral weight close to the valence band of the
insulating parent compound as the system is doped (for a review see
\cite{allen}).  This behavior is completely different from what one expects
in a simple rigid-band
picture in which the spectrum should be insensitive to doping with
only the Fermi energy changing. Obviously, a similar
trend is observed in the theoretical spectra where the quasiparticle resonance
due to the collective Kondo effect develops right at the top of the
lower Hubbard band as the system is doped (see Fig.~\ref{DOSdope}).

The quasi-two-dimensional nature of the cuprates is especially appealing
to the application of ARPES measurements. Accordingly the cuprates
have been exhaustively studied with this technique
\cite{campuzano,liu,king,olson,dessau}.
The general behavior found for the low-energy portion of the ARPES results
agrees with our results in that an experimental feature that might be
identified with the quasiparticle peak \cite{comment} crosses the
Fermi level in the $\Gamma$-$M$ direction just like the theoretical result
in Fig.~\ref{arpes}. Assuming the low-energy excitations to be
Fermi-liquid-like
\cite{liuand}, produces a quasiparticle bandstructure that also shows flat
bands near the $X$ point of the Brillouin-zone \cite{dessau,abrik,gofron}
identical to the theoretically generated quasiparticle band
in Fig.~\ref{bandstruct}. Since our results were obtained within a
dynamical MFT that is rather insensitive to details of the underlying lattice
structure, the agreement between  experiment and theory is strong evidence
that the low-energy single-particle dynamics are produced by strong
electronic correlations independent of the dimensionality!

\subsection{Optical conductivity.}
\begin{figure}[htb]
{\centerline{\psfig{figure=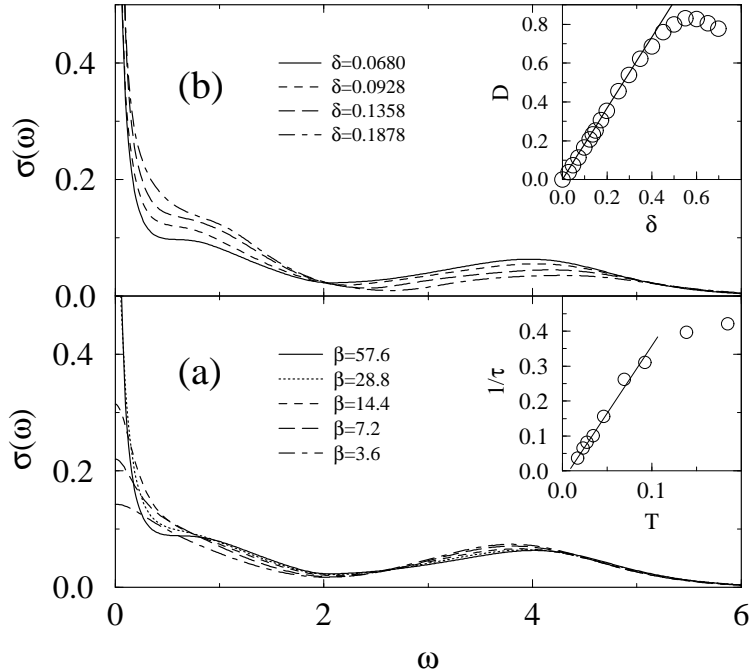,width=3.5in}}}
\caption[]{(a) {\em Optical conductivity vs. $\omega$ for various
temperatures when $\delta=0.068$.  Note that at low temperatures,
when the Kondo peak becomes pronounced in the DOS, a mid-IR feature begins
to appear in $\sigma(\omega)$.  As shown in the inset, if we fit this
data to a Drude form, then the width of the Drude peak is found to increase
roughly linearly with $T$.\\}
(b) {\em Filling dependence of the optical conductivity
when $U=4$ and $\beta=43.2$.  Note that for larger $\delta$ the mid-IR
and Drude peaks begin to merge, so that the latter is less distinct. The
inset shows the evolution of the Drude weight $D$ as a function of doping.
$D$ is determined by the extrapolation method of
Ref.~\cite{scal_white_zhang}. }}
\label{optical}
\end{figure}
\paragraph{Theory.}
The optical conductivity $\sigma(\omega)$
is another important probe of a strongly correlated
system.  It measures the rate at which electron-hole pairs are created
by photons of frequency $\omega$.
Figure~\ref{optical} shows the results \cite{pru_cox_ja,ja_free_pru}
for $\sigma(\omega )$
obtained from Eq. (\ref{optcond}) when $\delta=0.068$ for a variety of
temperatures (a), and results for $\beta=43.2$ and various dopings
$\delta$ (b). The Drude peak at $\omega=0$ develops with decreasing
temperature.
In addition there appears a small mid-infrared peak at $\omega\approx 1$.
As shown in Fig.~\ref{optical}b, this
peak is more pronounced for small $\delta$ but remains visible even at
larger doping. In addition, it is strongly temperature dependent and clearly
visible only for the lowest temperatures. We attribute it to excitations
from the lower Hubbard band to the quasiparticle band at the chemical
potential.  The last feature in $\sigma(\omega)$  is a roughly
temperature-independent peak at $\omega\approx U$  due to the charge
excitations from the lower part of the spectrum, i.e.\ from
the lower Hubbard band and the quasiparticle
peak at $\mu$, to the upper Hubbard band (cf.\ Fig.~\ref{DOStemp}).

The insets to Fig.~\ref{optical} show the development of the Drude weight as
obtained from Eq.\ (\ref{drude_weight}) as a function of doping
(Fig.~\ref{optical}b) and the width of the Drude peak as function of
temperature (Fig.~\ref{optical}a).  The latter was obtained by fitting the
generic form
\begin{equation}
\sigma(\omega\to 0)=\frac{D}{\pi}\frac{\tau}{1+\tau^2\omega^2}
\end{equation}
to the low-frequency regions in Fig.~\ref{optical}a.
The Drude weight initially increases linearly with $\delta$, then saturates
to its maximal value at $\delta\approx0.5$, before decreasing. This behavior
can be understood in terms of
a simple picture: The Drude weight is determined by the carrier density
and effective mass via $D\propto n/m^\ast$. From the doping-dependence
of the quasiparticle peak in the spectra in Fig.~\ref{DOSdope} one may assume
${m^\ast}^{-1}\propto\delta$.  The carrier density, on the other hand, is
given by $n\sim 1-\delta$, i.e.\ $D\propto\delta(1-\delta)$.
This expression explains the behavior for small doping as well as the
maximum, which should lie at $\delta_{max}\approx0.5$, and explains how the
character of the carriers changes from being holelike near half filling to
being electronlike at low density.

The width of the Drude peak, $1/\tau$, displays a linear behavior
$1/\tau\sim T$ for
$T\alt 0.1$.
This dependence may be traced back to the
development of the Kondo peak below $T_0$. Note that the intercept of the
linear
region does not lead to $1/\tau\to 0$ as $T\to 0$ as required, implying that
for very low temperatures the Drude width decreases more rapidly
($1/\tau\sim T^2$), as expected for a Fermi liquid.

%
\paragraph{Experiment.}
When one tries to make contact with experiment, care has to be taken about the
energy scales. The single-band Hubbard model should only be used to describe
low-energy features of the cuprates \cite{zr87}.
It does not make sense to use our results beyond $\omega\approx2$
in a comparison to experimental data, because the higher-energy bands,
corresponding to the charge-transfer insulating behavior of the parent
compounds, have been neglected in the mapping to a single-band Hubbard model.

Two types of experiments have been performed to determine the optical
conductivity of the cuprates: A Kramers-Kronig analysis was employed to
extract $\sigma(\omega )$ from reflectivity measurements
\cite{tanner_rev,thomas,uchida,tanner,renk}; and photoinduced absorption has
also been used \cite{heeger,kim}.  These experiments yield five low-energy
trends:
(i) The mid-IR peak maximum moves to lower frequency and merges with the Drude
peak as the doping increases; its spectral weight grows very rapidly with
doping; (ii) At a fixed value of doping, spectral weight rapidly moves to
lower $\omega$ as $T\rightarrow 0$, but the total weight in the Drude plus
mid-IR peaks remains approximately constant; the width of the Drude peak
decreases linearly with $T$; (iii) The insulating phase has a charge-transfer
gap; upon doping, the optical conductivity initially increases within
the gap region; (iv) There is an isobestic point, or nearly isobestic
behavior (in the sense that the optical conductivity is independent of
doping) at a frequency that is approximately one half the charge-transfer
gap; (v) more than one peak is observed in the mid-IR region.

Most of these trends are observed in the theoretical model:
the mid-IR peak is observed to move to lower frequency and merge with the
Drude peak as a function of doping; the optical conductivity rapidly increases
within the gap region as the system is doped; and there is an isobestic point
at
a frequency $\omega\approx 2$.  In addition, spectral weight is transferred
to lower frequencies as the temperature is decreased, and the Drude width
depends linearly on the temperature for a wide range of $T$, but the total
Drude plus mid-IR spectral weight increases as $T\rightarrow 0$.  The
theoretical model also does not display multiple mid-IR peaks.

These experimental features in the optical conductivity
are usually attributed to either phonons or impurities, but judging from our
results, the low-energy feature may also be due to excitations from the lower
Hubbard band to a dynamically generated quasiparticle band at the chemical
potential, i.e.\ connected to local or short-ranged spin-fluctuations.
Let us stress, though, that we do not want to make the point that one is able
to explain the normal-state of the cuprates in all respects {\em
quantitatively}
by studying
the single-band Hubbard model in the dynamical mean-field theory. We do think,
that our calculations display the underlying physics that drives the
anomalous features found in the normal state of the cuprates, and they
determine the order of magnitude of the corresponding temperature scale.

\subsection{Transport coefficients}
\begin{figure}[htb]
{\centerline{\psfig{figure=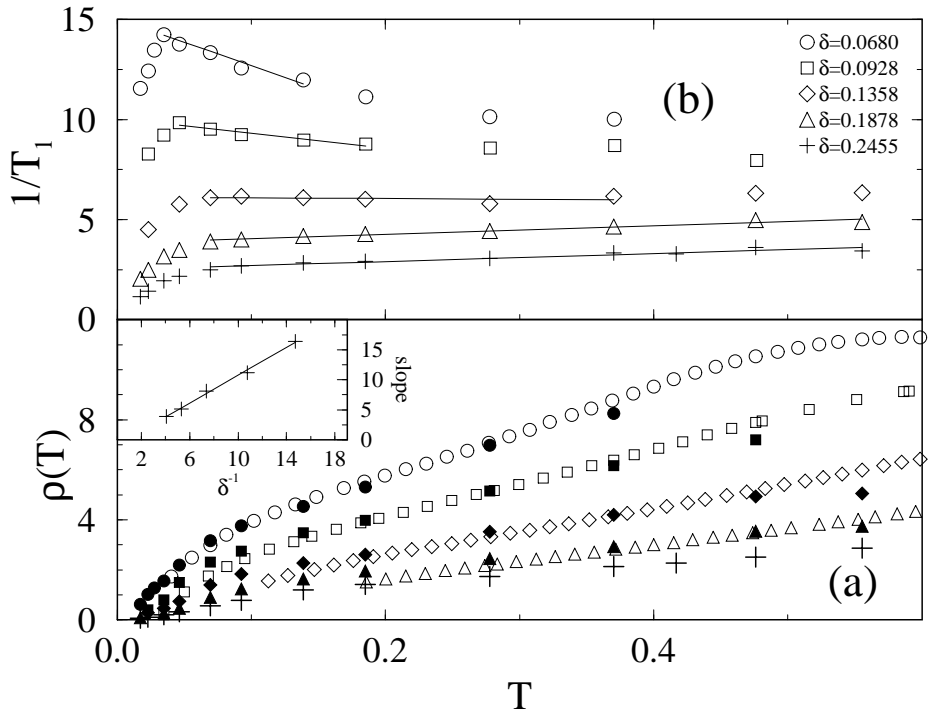,width=4.5in}}}

\caption[]{(a) {\em Resistivity versus temperature for several
different dopings when $U=4$.  The open (filled) symbols are for
NCA (QMC) results.  There is essentially exact agreement between the
NCA and QMC for high temperatures.  The slope in the
linear regime, determined by a linear least squares fit, increases
linearly with $1/\delta$ as shown in the inset. The units on the vertical axis
are $10^3\Omega cm$.}\\
(b) {\em NMR relaxation $1/T_1$ vs.\ temperature for different
dopings at $U=4t^*$.  The lines are linear fits in the anomalous region.}}
\label{resisnmr}
\end{figure}

\paragraph{Theory}
The creation of a dynamical low-energy scale, like the Kondo
temperature observed in the Hubbard model above,
is known to be accompanied by interesting and anomalous features
in transport coefficients. In our previous studies we already described
the resistivity and NMR relaxation rate as striking examples for this
behavior.  Figure~\ref{resisnmr} summarizes these results.
Most prominent is the pronounced linear region in $\rho(T)$
which increases with increased doping. The slope of this region is
proportional to $1/\delta$ as shown in the inset to Fig.~\ref{resisnmr}a.
The thermal conductivity $\kappa$ was also calculated; however it is not
plotted, since to a very good approximation, it follows the Wiedemann-Franz law
$\kappa\propto T/\rho$.  Figure~\ref{resisnmr}b presents the results for the
NMR relaxation rate ($T_1$). Here, too, a rather anomalous
variation with both temperature and doping is found. A linear
region in $1/T_1$ develops as the doping increases, and the slope changes
sign for $\delta\approx 0.15$.  The doping dependence of $1/T_1$ is reduced
as the temperature increases, but does not disappear at the temperatures that
can be reached by the numerical analytic continuation.

\begin{figure}[htb]
{\centerline{\psfig{figure=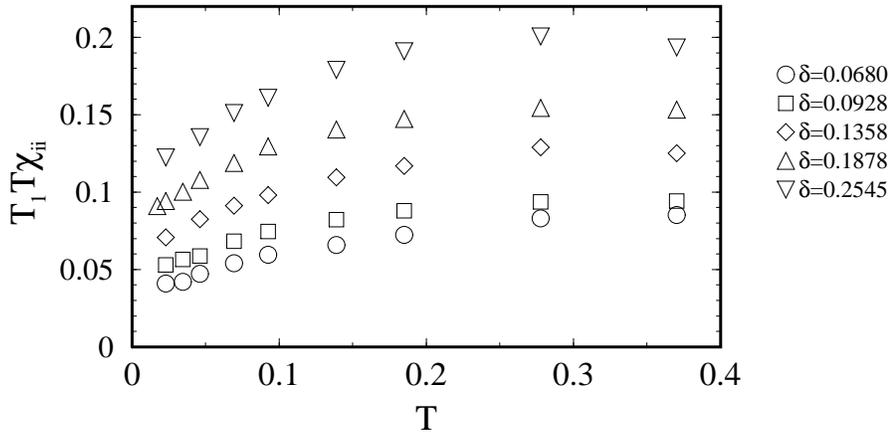,width=4.5in}}}
\caption[]{{\em Ratio of the longitudinal $(T_1)$ and transverse
$(T_{2g}\propto 1/\chi_{loc})$ NMR-relaxation
rates multiplied by $T$.  Note how this ratio becomes flat at intermediate
values of $T$, which is supposed to be a signal for the quantum-critical
regime.  The constant value of this ratio increases with doping, but is
nearly constant near half-filling.}}
\label{qcrit}
\end{figure}

It has been recently argued that the constancy of the ratio $T_1T/T_{2G}$ is
a test for the quantum critical region of the two-dimensional Heisenberg spins
in the Cu-O planes of the cuprates \cite{sokol} (even though this theory
neglects the charge degrees of freedom, assuming that the most important
effect of the holes is to add {\em disorder} into the
spin system).  However, as seen in Fig.~\ref{qcrit}, this ratio also becomes
flat at intermediate temperatures within the dynamical mean-field theory in
infinite dimensions.  The constant value of the ratio $T_1T/T_{2G}$ increases
with doping, but does not change significantly for the lowest values of the
doping.

\begin{figure}[htb]
{\centerline{\psfig{figure=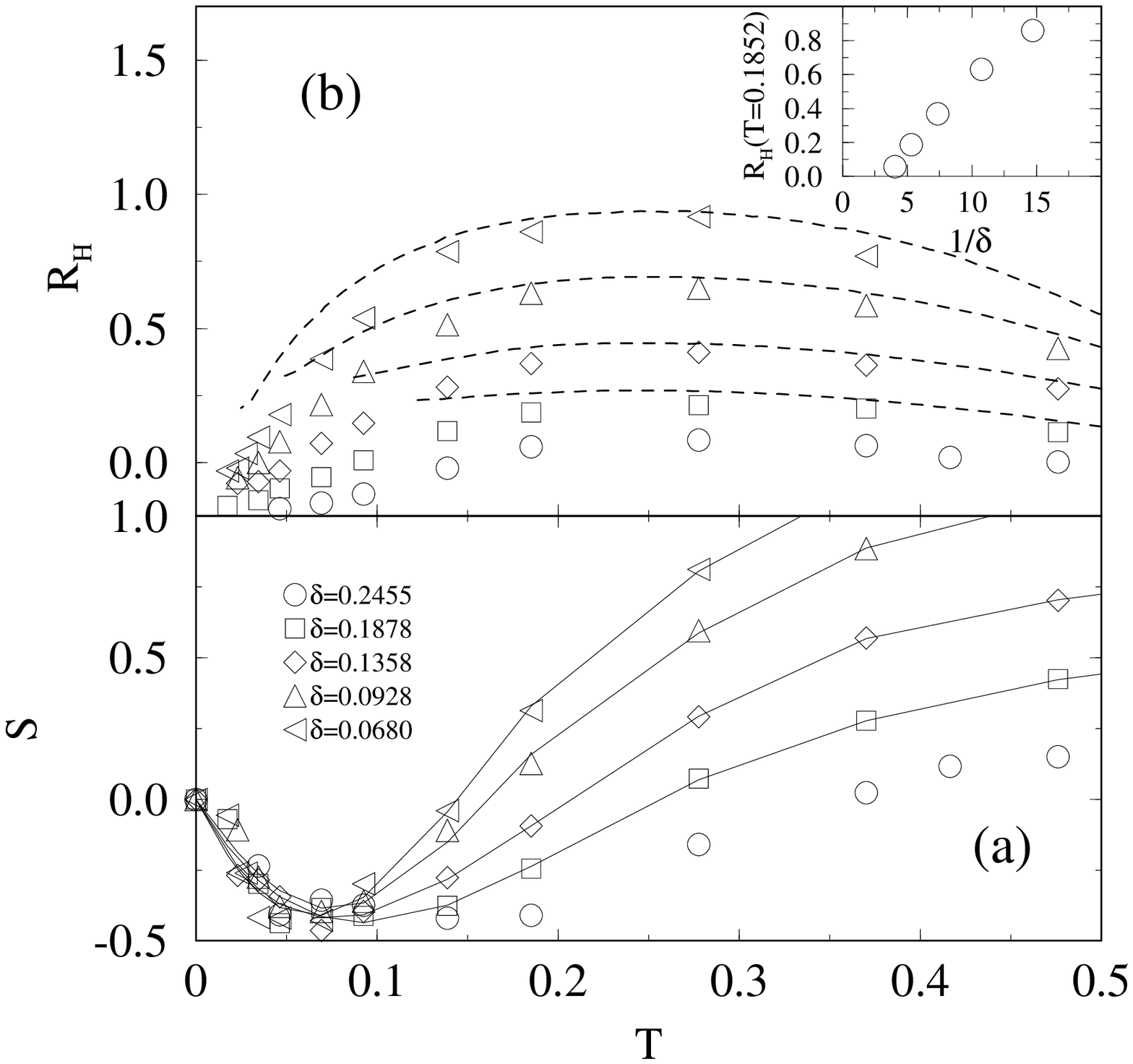,width=4.5in}}}
\caption[]{{\em Thermopower (a) and Hall coefficient (b) for four different
dopings $\delta$ as function of temperature.  In (b), the open symbols
(dashed lines) are QMC (NCA) results and in (a) the solid lines come from
a fit to the QMC data (open symbols).  In the inset to (a), the
Hall coefficient at a fixed temperature $R_H(T=0.1852)$ is plotted versus
$1/\delta$, indicating that $R_H(T=0.1852)$ increases roughly in proportion
to $1/\delta$, consistent with experimental results for the
cuprates\cite{hall_exp}. The units on the vertical axis are $86\mu V/K$ for the
thermopower (a)
and $10^{-9} m^3/C$ for $R_H$ (b).}}
\label{thermohall}
\end{figure}
\begin{figure}[htb]
{\centerline{\psfig{figure=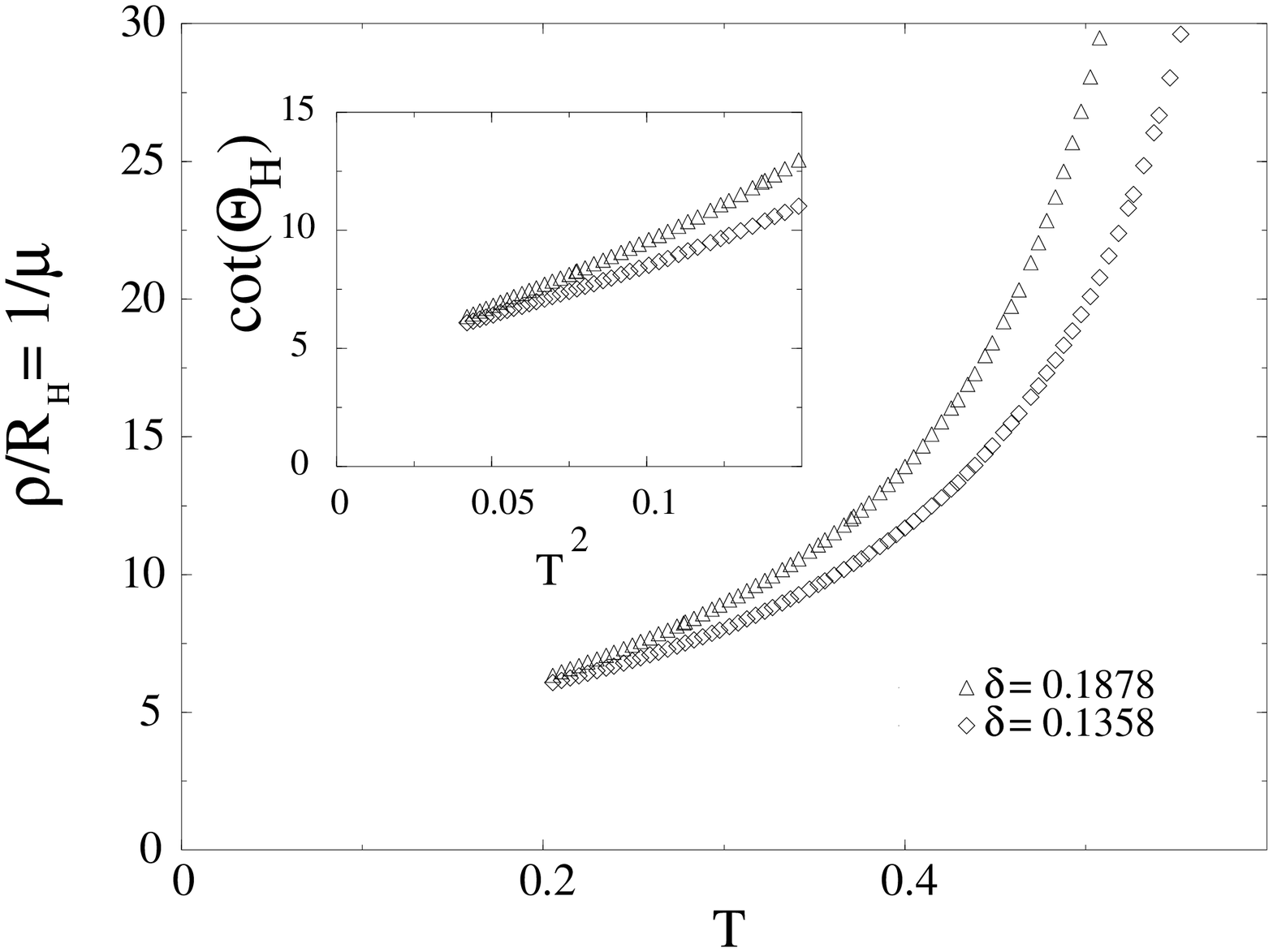,width=4.5in}}}

\caption[]{\em Inverse mobility $1/\mu = \rho/R_H$ calculated with
the NCA for two dopings as function of $T$ for large $T$.   The
NCA was used since it is essentially exact at these temperatures
(cf.\ Fig.~\ref{thermohall}), and significantly less computationally
expensive, thus allowing us to present a dense data set.  In the inset
the Hall angle $\cot\Theta_H$ is plotted as function
of $T^2$. Note the apparent linear behavior.}
\label{hall_angle}
\end{figure}
%
Finally we want to report calculations of two other interesting transport
coefficients, namely the thermopower $(S)$ and the Hall coefficient ($R_H$).
The strange behavior of the latter gave rise to a number of speculations
about different scattering mechanisms for transport with and without
magnetic field \cite{and91} and it is thus interesting to look at the
behavior of this quantity within our scheme. Fig.~\ref{thermohall} compiles
the results for $S$ (Fig.~\ref{thermohall}a) and $R_H$ (Fig.~\ref{thermohall}b)
for four different dopings as function
of temperature.  The thermopower shows the sign change at intermediate
temperatures that is
characteristic of correlated materials. The Hall coefficient,
in Fig.~\ref{thermohall}b, is
more interesting.   It is positive for high temperatures,  and displays
a maximum at intermediate temperatures, followed by a strong
decrease for lower $T$, eventually becoming negative (qualitatively
similar features have been seen in simulations of the two-dimensional Hubbard
model \cite{assaad}). Interestingly, the position of
the maximum is weakly sensitive to doping while, as shown in the inset,
its height roughly decreases with increasing doping like $R_H\sim 1/\delta$.
Finally, we show the quadratic behavior of the Hall angle
($\cot\Theta_H=\rho_{xx}/
\rho_{xy}=1/\mu$, where $\mu$ is the charge carrier's mobility).
The Hall angle is calculated with the NCA so that a dense set of points
may be presented at high temperatures, where the analytic continuation
of QMC data becomes numerically expensive.  Here the agreement between
the NCA and the QMC data is essentially exact (cf.\ Fig.~\ref{thermohall}).
Fig.~\ref{hall_angle} plots
the inverse mobility as function of temperature for two dopings (c.f.\
Fig.~\ref{experiments}
for typical experimental temperature and doping dependence in the cuprates).
The Hall angle is plotted
as a function of $T^2$ in the inset to Fig.~\ref{hall_angle}.
It shows a roughly linear behavior on this scale.

\paragraph{Experiment.}
As we have pointed out \cite{pru_ja},
the theoretical behavior is consistent with the normal-state
properties of hole-doped high-$T_c$ compounds. As is well known, the
experimental
trends observed include \cite{rew_oxides,hall_exp} (cf.\ also
Fig.~\ref{experiments}):
(i) A (sub-) linear variation of the resistivity with temperature, the slope
or absolute value of $\rho(T)$ for a given $T$ decreasing with $1/\delta$;
(ii) The NMR-rate drops with increasing doping due to a decrease of the
(local) spin fluctuations and shows a linear tail with positive slope at
intermediate
temperatures; There is also a tendency towards a change in sign of the slope
of this linear region with decreasing doping;
(iii) The spin-lattice relaxation rate $T_1$ becomes doping independent at
intermediate values of the temperature, and the ratio $T_1T/T_{2G}$ becomes
constant over a similar temperature range;
(iv) The Hall coefficient goes through a maximum whose position is roughly
independent of doping but whose height decreases $\propto 1/\delta$; (v)
The Hall angle $\cot\Theta_H$ is usually found to vary
like $T^2$ over a considerable temperature range.
Since all these features (except the constancy of $1/T_1$ with respect to
doping
at intermediate $T$) are consistently obtained from the dynamical MFT,
including the peculiar behavior of the Hall angle (without
any necessity to resort to an exotic ground-state or scattering mechanism) and
the constancy of the ratio $T_1T/T_{2G}$,
we anticipate that these anomalies are intrinsic properties of
strong local correlations and do not specifically
depend on the low dimensionality
of the cuprates. This statement is further supported by the observation that
similar features in the NMR and EPR relaxation rates and the Hall coefficient
are also found in ``conventional''
heavy-fermion materials \cite{gresteg,loidl}.


\section{Summary and conclusion}
We presented results for a variety of transport properties that support
the conjecture that most of the anomalous normal-state properties of
the high-$T_c$ compounds are intrinsic properties resulting from correlation
effects in a single-band model. These anomalies include a linear in $T$
resistivity and anomalies in the temperature dependence of the NMR-relaxation
rate and
thermopower. We also presented a method to calculate the Hall-conductivity and
Hall coefficient that does not rely on the (conventionally used)
relaxation-time approximation or the introduction of a new scattering
mechanism like Coleman's ``skew-scattering''.
This new method enabled us to obtain physically sensible
results for the Hall coefficient, showing a positive sign and a temperature
dependence characteristic of strongly correlated systems. We also found that
the
Hall angle or inverse mobility of the carriers shows a clear $T^2$-behavior
as also observed in the cuprates. The optical conductivity was found
to have a Drude peak at small $\omega$, a charge-excitation peak
at $\omega\approx U$, and a mid-IR feature that is attributed to excitations
from the lower Hubbard band to the quasiparticle band dynamically generated at
the Fermi level.

Most of these properties also show a distinctive dependence upon doping: The
Drude weight
increases linearly with doping $\delta$ and both the slope of the linear region
in the resistivity and the Hall coefficient increase $\sim 1/\delta$. The
mid-IR
feature in the optical conductivity initially increases rapidly with doping,
and then decreases as $\delta$ increases further (but is still
visible up to $\delta\approx 0.20$).

These features are also found in the high-$T_c$ materials and to some extent
in heavy-fermion or mixed-valence compounds.
{}From our results it seems that
the appearance and overall temperature dependence of these anomalies
is due to the existence of strong local correlations that
lead to a dynamically generated (strongly temperature-dependent)
low-energy scale arising from a Kondo-like screening of the (local)
magnetic moments. The anomalous regions are more pronounced in the present
case, which may be attributed to the fact that in the Hubbard model
only one band exists, i.e.\ the electrons that form the moments are
also responsible for their screening, while these different tasks are split
between
at least two bands in the periodic Anderson model. To distinguish between
these different physical situations we suggest that the singlet formation
found in the Hubbard model should be labeled as a collective
``single-band Kondo-effect''.

Motivated by these considerations, we propose that the peculiar anomalies found
in the
cuprates may be viewed as a crossover phenomenon from a high-temperature
``normal'' phase to a renormalized Fermi liquid as $T\to 0$. The anomalous
temperature dependence of physical quantities is then obtained from the
peculiar $T$-behavior of the
developing quasiparticle band at $\mu$. Note that in several heavy-fermion
compounds true Fermi-liquid behavior is not observed, because the development
of
the Fermi liquid is preempted by a phase transition into an ordered state. It
was fortuitous that the first heavy-fermion compounds studied did have
transition temperatures much smaller than $T_K$, enabling one to observe
the formation of the heavy Fermi liquid first.
In the cuprates, on the other hand, it may be that the
relevant Kondo scale is of the same order as the transition temperatures,
and we are in a situation where the mentioned crossover is observed,
but that the system undergoes a phase transition before
a Fermi liquid forms \cite{levin}.

\paragraph{Acknowledgments}  We would like to acknowledge useful
conversations with
F.F.\ Assaad,
W.\ Chung,
J.\ Keller,
Y.\ Kim,
C.\ Quitmann,
D.\ Scalapino,
R. Scalettar,
R. Singh,
D.\ Tanner,
and G.\ Thomas.
This work was supported by the National Science Foundation grant number
DMR-9107563, the
NATO Collaborative Research Grant number CRG 931429 and
through the NSF NYI program.  In addition, we
would like to thank the Ohio Supercomputing Center, and the Physics
department of the Ohio State University for providing computer facilities.


\begin{thebibliography}{99}
\bibitem{comment} The assumption of this peak being Fermi-liquid like is still
heavily discussed. See e.g.\ the work in \cite{olson} and \cite{liuand}.
\end{thebibliography}
\end{document}